# 1.064-μm-band up-conversion single-photon detector

Fei Ma,[1] Ming-Yang Zheng,[2] Quan Yao,[3] Xiuping Xie,[2] Qiang Zhang,[1, 3, *] and Jian-Wei Pan[1]

[1]*Hefei National Laboratory for Physical Sciences at Microscale and Department of Modern Physics, University of Science and Technology of China, Hefei, Anhui 230026, China*

[2]*Shandong Institute of Quantum Science and Technology Co., Ltd., Jinan, Shandong 250101, China*

[3]*Jinan Institute of Quantum Technology, Jinan, Shandong 250101, China*

*\*qiangzh@ustc.edu.cn*

## 1. Abstract

Based on the technique of periodically poled lithium niobate (PPLN) waveguide, up-conversion single-photon detection at 1.064-μm is demonstrated. We have achieved a system photon detection efficiency (DE) of 32.5% with a very low noise count rate (NCR) of 45 counts per second (cps) by pumping with a 1.55-μm-band single frequency laser using the long-wavelength pumping technique and exploiting volume Bragg grating (VBG) as a narrow band filter. Replacing the VBG with a combination of adequate dielectric filters, a DE of up to 38% with a NCR of 700 cps is achieved, making the overall system more stable and practical. The up-conversion single-photon detector (SPD) operating at 1.064 μm can be a promising robust counter and find usage in many fields.

## 2. Introduction

Various SPDs working at different spectral windows have been developed, including silicon avalanche photodiode (APD) [1-3], InGaAs/InP APDs [4, 5], up-conversion SPD [6, 7], and superconducting SPD [8, 9]. SPDs have extensive applications in quantum communication systems [10], optical time domain reflectometry (OTDR) [11, 12], single-photon-level spectrometer [13, 14], high resolution imaging in astronomy [15] and so forth.

On the other hand, 1.064-μm band is useful in a wide range of applications, such as lidar [16], deep space optical communication [17, 18] and biomedical spectroscopy [19], attributing to the corresponding commercial lasers with many advantages of high power, well-developed technology and compact configuration. Generally, the above applications are always in need of sensitive receivers since the back-scattered light is extremely weak. Therefore, SPDs operating at 1.064-μm band are much in demand. However, the 1.064-μm band is an awkward detection band because it is at the low-response regime of both silicon and InGaAs/InP APDs. The bandgap of silicon is 1.12 eV, making the absorption and the photo-response curve decrease precipitously for wavelength around 1 μm, while InGaAs/InP APDs are generally designed for telecom band near 1.55 μm. All the traditional commercial detectors are reported with a DE of < 5% at 1.064 μm [20, 21].

Due to the importance and necessity of single-photon detection at the 1.064-μm band, the enhanced InGaAs/InP APDs are invented. A DE of 10% and a NCR of 1000 cps are achieved in experiments [22]. ID400 advanced single-photon detection system, which is a commercial product, has a DE of 30% with a NCR of 2000 cps [23]. Although the enhanced InGaAs/InP APDs have good performance, the dead time of microsecond level limits the count rate.

The superconducting SPD has impressive performance with a DE of up to 40%, while the NCR is up to 27000 cps [24]. Moreover, the need for cryogenic cooling and exquisite temperature control may limit its wide use [8, 9].

Alternatively, up-conversion SPD is a promising technique to extend the detection range of well-developed silicon APD to near-infrared band. In literature, using bulk crystal PPLN device and 1.55-μm pulse pump, an up-conversion system at 1.06-μm band was demonstrated with a NCR of 150 cps at the expense of DE [25]. We can improve the performance of up-conversion SPD at 1.06-μm band by using PPLN waveguide devices.

Room-temperature waveguide-based up-conversion SPD for the communication wavelength signal around 1550 nm can have a DE higher than 40% [26], a NCR as low as 100 cps [27], and a dead time of nanosecond level [28]. The key for high performance up-conversion SPD is to obtain ultra-low noise. The noises in waveguide devices mainly come from the spontaneous parametric down-conversion (SPDC), spontaneous Raman scattering (SRS), parasitic noises caused by imperfect periodic poling structure and second and third harmonic generation (SHG and THG) when the strong pump goes through the waveguide. The long-wavelength pumping technique can eliminate the noise caused by SPDC and dramatically reduce the noise caused by SRS [29]. By using narrow-band filters, the SHG and THG of the pump can be blocked, the remnant SRS noise and parasitic noises can also be controlled because the NCR is proportional to the bandwidth of the filters [27, 30].

In our work, the signal photons at 1.064 μm are launched into a reverse-proton-exchanged (RPE) PPLN waveguide, pumped by a 1.55-μm single frequency laser and up-converted to photons in the visible band that are then detected with silicon APD. Using VBG as a narrow-band filter, an up-conversion SPD working at free-running mode has showed an ultra-low NCR of 45 cps.

## 3. Waveguide characterization and experimental setup

We fabricate RPE PPLN waveguides [31] for sum-frequency generation (SFG) of 1.55-μm pump and 1.064-μm signal. The total length of the waveguides is 52 mm and the length of the quasi-phase-matching (QPM) gratings with a QPM period of 10.2 μm is 48 mm. The mode filter at the input port of the waveguides is designed to maximize the coupling with the HI1060 optical fiber, and the fiber-pigtailing loss is ~1 dB. The input and output end faces of the waveguides are anti-reflection (AR) coated for both the signal and the sum-frequency wavelengths to eliminate the Fresnel reflection loss. The waveguides have a propagation loss of < 0.2 dB/cm at 1.55 μm measured with the Fabry-Perot fringe-contrast method [32].

The SFG tuning curve for a typical waveguide is shown in Fig. 1(a). The signal wavelength is fixed at 1.064 μm and the pump wavelength is swept around 1550 nm. The phase-matching wavelength is 1546.7 nm at room temperature with a full width at half maximum (FWHM) of 0.5 nm. The SFG photon conversion efficiency as a function of pump power [33] is shown in Fig. 1(b). The maximum conversion efficiency is obtained when the pump power at the input port of the PPLN waveguide is 85 mW, corresponding to a normalized conversion efficiency [26] of $126\%/(W \bullet cm^2)$.

A schematic diagram of our experimental setup is shown in Fig. 2.

A continuous-wave (CW), tunable external cavity diode laser (ECDL) is utilized as the pump seed source. The seed is amplified by an erbium-doped fiber amplifier (EDFA) which produces a maximum power of 200 mW at ~1.55 μm. Any unwanted photons at 1.064 μm and 0.631 μm generated in the EDFA are removed using a 1.064-μm/1.55-μm wavelength division multiplexer (WDM) with isolation of ~35 dB and a 0.631-μm/1.55-μm WDM with isolation of ~20 dB.

The 1.064-μm signal, which is one-million photons at the input port of the WDM with a power of -97.28 dBm, is provided by a single-frequency CW laser together with two variable optical attenuators [28]. A 1/99 beam splitter and a calibrated power meter are employed to monitor the input signal power.

As RPE waveguides support only TM-polarized modes, polarization controllers are used to adjust the polarization of both the signal and the pump beams. The single photon source is combined with the pump in a 1.064-μm/1.55-μm WDM whose output is connected to the fiber-pigtailed PPLN waveguide. The working temperature of the waveguide is kept at 38 ℃ by a thermoelectric cooling (TEC) system to maintain the phase-matching condition.

The up-converted photons and the remnant pump are collected by an AR-coated aspheric lens (AL), and then the remnant pump is removed by a dichroic mirror (DM).

The VBG-filtering scheme is shown in Fig. 2(b). A VBG with 85% reflection efficiency and 0.05-nm bandwidth in the 631-nm band can dramatically reduce the SRS noise and parasitic noises. A 631-nm band-pass filter (BPF) with 2.4-nm bandwidth is employed to block the SHG and THG of the pump and further suppress the parasitic noises. Then, the SFG photons are collected by an AR-coated AL. VBG-based up-conversion SPD has high performance while it may not be practical in field because VBG has critical position and angular tolerances, raising stability issues [27, 28].

The BPFs-filtering scheme is shown in Fig. 2(c). Four 631-nm BPFs are used to filter the noise photons generated by the strong pump, including the SHG and THG of the pump and the parasitic noises. Then, a 631-nm collimator and a piece of multimode fiber are used to collect the SFG photons. By optimizing the filtering system in this scheme with a special coating design, BPFs-based up-conversion SPD can form a robust all-fiber system and is easy to integrate [28].

Finally, the up-converted photons are detected by a silicon APD, which has a DE of approximately 70% at 631 nm and an intrinsic noise count rate of 20 cps.

## 4. Up-conversion single-photon detection performance

We tune the pump power and record the DE and NCR accordingly. The system DE is obtained by dividing the number of detected counts after NCR subtraction by one million, which is the signal photon count before entering the WDM. Curves of DE and NCR versus pump power for the two different filtering schemes are shown in Fig. 3.

VBG has a filtering bandwidth much narrower than that of BPFs, therefore the VBG-based up-conversion SPD has an ultra-low NCR of 45 cps for the maximum DE of 32.5%. The greatest strength of our SPD is its low noise-equivalent-power (NEP) [26]. When the NCR is reduced to the silicon APD's intrinsic count rate of 20 cps by lowering the pump power, the DE is still 26.5%, corresponding to a NEP of -143.5 $dBm/\sqrt{Hz}$.

The ultra-low NCR of the VBG-based up-conversion SPD benefits from the great signal-pump frequency difference of 2946.9 $cm^{-1}$ and the narrow band filters. Large signal-pump frequency difference means very weak anti-Stokes SRS noise [29]. The narrow band filters can reduce the nonlinear noise efficiently without sacrificing DE. The noises from SHG and THG of the pump are removed by the BPF. In addition, by the narrow-band filtering of the VBG with a bandwidth of 0.05-nm at the 631-nm band, SRS noise and parasitic noises are greatly eliminated.

The BPFs-filtering scheme allows a higher DE at the cost of a higher NCR. The BPFs-filtering scheme has a maximum DE of 38% with a NCR of 700 cps, i.e. a NEP of -137.4 $dBm/\sqrt{Hz}$. The filtering system in this scheme can be integrated as a fiber filter and make the overall system more stable and practical in field [28].

The measured DE is consistent with our estimations by combining the throughput of each component, as shown in Table 1. Combining with the silicon APD's DE of ~70%, the theoretical DE of the VBG and BPFs filtering schemes are consistent with our experimental results.

**Table 1. Loss and throughput of the up-conversion SPD components**

| Component | Loss(dB) | Throughput |
|---|---|---|
| WDM | 0.3 | 93% |
| Waveguide | 1.9 | 65% |
| VBG filtering system | 1.14 | 77% |
| BPFs filtering system | 0.45 | 90% |

## 5. Conclusion

We have demonstrated an up-conversion detector for the 1.064-μm band using RPE PPLN waveguide. Two kinds of filtering schemes are investigated, achieving either ultra-low noise and adequately high DE for high-standard experimental demand or high DE with adequately low NCR for field use. It is conceivable that our up-conversion SPD at 1.064-μm has potential applications in many fields, such as single photon imaging, atmospheric research, and medical devices. In the following research, the 1.064-μm single photon detector would be applied in photon radar experiments for atmosphere visibility and wind speed detection.

**Acknowledgments**

This work has been supported by the National Fundamental Research Program (under Grant No. 2013CB336800), the National Natural Science Foundation of China, the Chinese Academy of Science, SAICT Experts Program, and the 10000-Plan of Shandong Province.

**References**

1. H. Dautet, P. Deschamps, B. Dion, A. D. MacGregor, D. MacSween, R. J. McIntyre, C. Trottier, and P. P. Webb, "Photon counting techniques with silicon avalanche photodiodes," Appl. Opt. **32**(21), 3894–3900 (1993).
2. S. Cova, M. Ghioni, A. Lacaita, C. Samori, and F. Zappa, "Avalanche photodiodes and quenching circuits for single-photon detection," Appl. Opt. **35**(12), 1956–1976 (1996).
3. M. Ghioni, A. Gulinatti, I. Rech, F. Zappa, S. Cova, "Progress in Silicon Single-Photon Avalanche Diodes," IEEE J. Sel. Top. Quantum Electron. **13**(4), 852–862 (2007).
4. A. Lacaita, F. Zappa, S. Cova, and P. Lovati, "Single-photon detection beyond 1 μm: performance of commercially available InGaAs/InP detectors," Appl. Opt. **35**(16), 298–2996 (1996).
5. A. Tosi, A. D. Mora, F. Zappa, and S. Cova, "Single-photon avalanche diodes for the near-infrared range: detector and circuit issues," J. Mod. Opt. **56**(2-3), 299–308 (2009).
6. P. Kumar, "Quantum frequency conversion," Opt. Lett. **15**(24), 1476–1478 (1990).
7. A. P. Vandevender and P. G. Kwiat, "High efficiency single photon detection via frequency up-conversion," J. Mod. Opt. **51**(9-10), 1433–1445 (2004).
8. B. Cabrera, R. M. Clarke, P. Colling, A. J. Miller, S. Nam, and R. W. Romani, "Detection of single infrared, optical, and ultraviolet photons using superconducting transition edge sensors," Appl. Phys. Lett. **73**(6), 735–737 (1998).
9. G. N. Gol'tsman, O. Okunev, G. Chulkova, A. Lipatov, A. Semenov, K. Smirnov, B. Voronov, A. Dzardanov, C. Williams, and R. Sobolewski, "Picosecond superconducting single-photon optical detector," Appl. Phys. Lett. **79**(6), 705–707 (2001).
10. R. H. Hadfield, "Single-photon detectors for optical quantum information applications," Nat. Photonics **3**(12), 696–705 (2009).
11. B. F. Levine, C. G. Bethea, L. G. Cohen, J. C. Campbell, and G. D. Morris, "Optical time domain reflectometry using a photon-counting InGaAs/InP avalanche photodiode at 1.3 μm," Electron. Lett. **21**(2), 83–84 (1985).
12. M. Legré, R. Thew, H. Zbinden, and N. Gisin, "High resolution optical time domain reflectometer based on 1.55 μm up-conversion photon-counting module," Opt. Express **15**(13), 8237–8242 (2007).
13. Q. Zhang, C. Langrock, M. M. Fejer, and Y. Yamamoto, "Waveguide-based single-pixel up-conversion infrared spectrometer," Opt. Express **16**(24), 19557–19561 (2008).
14. L. Ma, O. Slattery, and X. Tang, "Experimental study of high sensitivity infrared spectrometer with waveguidebased up-conversion detector," Opt. Express **17**(16), 14395–14404 (2009).
15. J.-T. Gomes, L. Delage, R. Baudoin, L. Grossard, L. Bouyeron, D. Ceus, F. Reynaud, H. Herrmann, and W. Sohler, "Laboratory demonstration of spatial-coherence analysis of a blackbody through an up-conversion interferometer," Phys. Rev. Lett. **112**(14), 143904 (2014).
16. M. Pfennigbauer and A. Ullrich, "Applicability of single photon detection for laser radar," e & i Elektrotechnik und Informationstechnik **124**(6), 180–185 (2007).
17. K. Wilson and M. Enoch, "Optical communications for deep space missions," IEEE Commun. Mag. **38**(8), 134–139 (2000).
18. S. Vasile, M. S. Ünlü, J. Lipson, "Challenges of developing resonant cavity photon-counting detectors at 1064 nm," in *Free-Space Laser Communication Technologies XXII*, H. Hemmati, ed., Proc. SPIE **7587**, 75870T-2 (2010).
19. H. Yamazaki, S. Kaminaka, E. Kohda, M. Mukai, and H. Hamaguchi, "The diagnosis of lung cancer using 1064-nm excited near-infrared multichannel Raman spectroscopy," Radiat. Med. **21**(1), 1–6 (2003).
20. Excelitas Technologies, Canada, http://www.excelitas.com/Downloads/DTS_SPCM-AQRH.pdf.
21. ID Quantique, "Infrared single-photon counter," http://www.photonicsolutions.co.uk/upfiles/id230-specs.pdf.
22. M. A. Itzler, X. D. Jiang, R. Ben-Michael, K. Slomkowski, M. A. Krainak, S. Wu, and X. L. Sun, "InGaAsP/InP avalanche photodetectors for non-gated 1.06 μm photon-counting receivers," in *Enabling Photonics Technologies for Defense,*

*Security, and Aerospace Applications III*, M. J. Hayduk, A. R. Pirich, P. J. Delfyett Jr., E. J. Donkor, J. P. Barrios, R. J. Bussjager, M. L. Fanto, R. L. Kaminski, G. Li, H. Mohseni, and E. W. Taylor, eds., Proc. SPIE **6572**, 65720G (2007).
23. ID Quantique, "1064nm single-photon counter," http://www.photonicsolutions.co.uk/upfiles/id400-specs.pdf.
24. J. A. Stern and W. H. Farr, "Fabrication and characterization of superconducting NbN nanowire single photon detectors," IEEE Trans. Appl. Supercond. **17**(2), 306–309 (2007).
25. H. Dong, H. Pan, L. Yao, E. Wu, and H. Zeng, "Efficient single-photon frequency upconversion at 1.06 μm with ultralow background counts," Appl. Phys. Lett. **93**(7), 071101 (2008).
26. C. Langrock, E. Diamanti, R. V. Roussev, Y. Yamamoto, M. M. Fejer, and H. Takesue, "Highly efficient single-photon detection at communication wavelengths by use of upconversion in reverse-proton-exchanged periodically poled LiNbO$_3$ waveguides," Opt. Lett. **30**(13), 1725–1727 (2005).
27. G. Shentu, J. S. Pelc, X. Wang, Q. Sun, M. Zheng, M. M. Fejer, Q. Zhang, and J. Pan, "Ultralow noise up-conversion detector and spectrometer for the telecom band," Opt. Express **21**(12), 13986–13991 (2013).
28. M. Zheng, G. Shentu, F. Ma, F. Zhou, H. Zhang, Y. Dai, X. Xie, Q. Zhang, and J. Pan, "Integrated four-channel all-fiber up-conversion single-photon-detector with adjustable efficiency and dark count," Rev. Sci. Instrum. **87**(9), 093115 (2016).
29. J. S. Pelc, L. Ma, C. R. Phillips, Q. Zhang, C. Langrock, O. Slattery, X. Tang, and M. M. Fejer, "Long-wavelength-pumped upconversion single-photon detector at 1550 nm: performance and noise analysis," Opt. Express **19**(22), 21445–21456 (2011).
30. P. S. Kuo, J. S. Pelc, O. Slattery, Y. Kim, M. M. Fejer, and X. Tang, "Reducing noise in single-photon-level frequency conversion," Opt. Lett. **38**(8), 1310–1312 (2013).
31. K. R. Parameswaran, R. K. Route, J. R. Kurz, R. V. Roussev, M. M. Fejer, and M. Fujimura, "Highly efficient second-harmonic generation in buried waveguides formed by annealed and reverse proton exchange in periodically poled lithium niobate," Opt. Lett. **27**(3), 179–181 (2002).
32. R. Regener and W. Sohler, "Loss in low-finesse Ti:LiNbO$_3$ optical waveguide resonators," Appl. Phys. B **36**(3), 143–147 (1985).
33. All pump power mentioned below is measured at the input port of the PPLN waveguide.

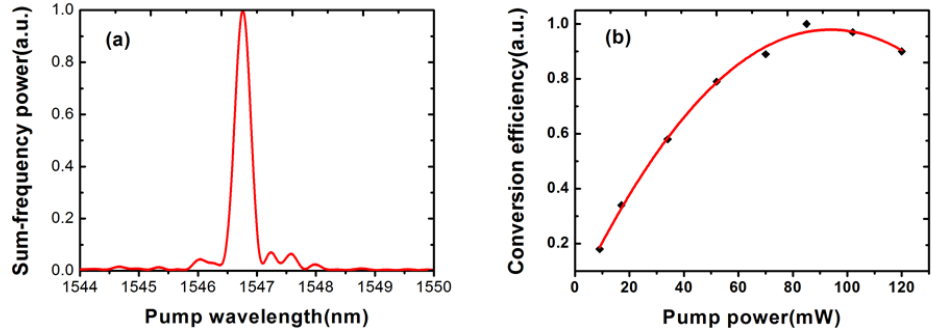

Fig. 1. (a) Tuning curve; (b) Measured SFG photon conversion efficiency (black diamonds) and theoretical fit (red line).

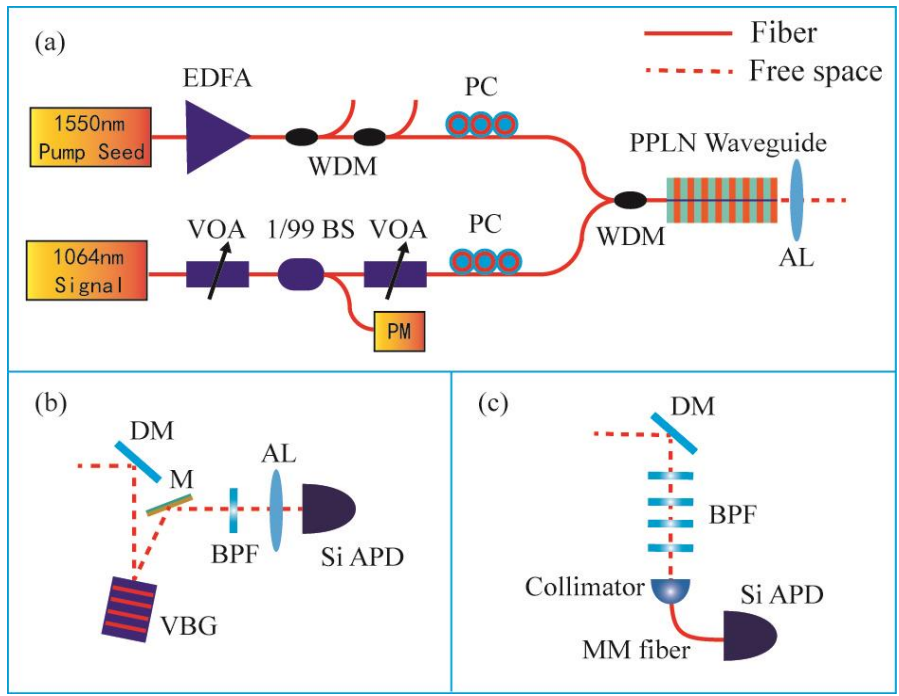

Fig. 2. (a) Schematic diagram of the up-conversion SPD at 1.064 um. The waveguide output is filtered by (b) a VBG and a BPF or (c) four BPFs. EDFA, erbium-doped fiber amplifier; WDM, wavelength division multiplexer; PC, polarization controller; VOA, variable optical attenuator; BS, beam splitter; PM, power meter; AL, aspheric lens; DM, dichroic mirror; VBG, volume Bragg grating; M, mirror; BPF, band-pass filter; Si APD, silicon avalanche photodiode; MM fiber, multimode fiber.

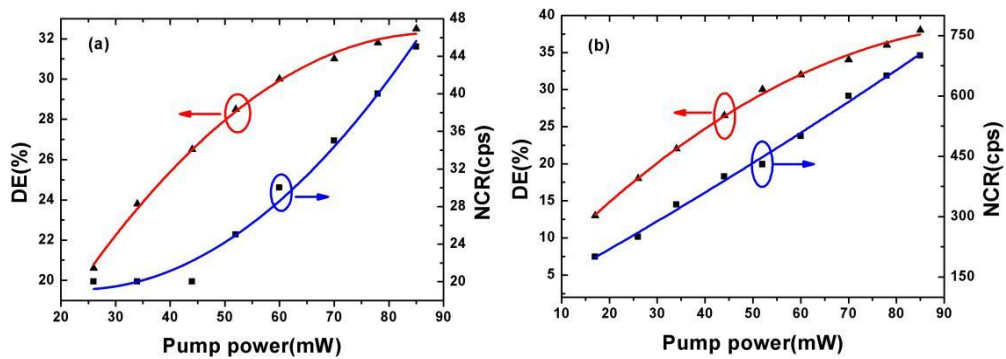

Fig.3. DE (red line, triangle) and NCR (blue line, square) versus pump power with two different filtering schemes. (a) VBG filtering, (b) BPFs filtering.